\documentstyle[12pt]{article}

\begin{document}

\title{\bf Explicit Modular Invariant \\Two-Loop Superstring Amplitude\\
           Relevant for $R^4$}

\author{Roberto Iengo\thanks{E-mail: iengo@he.sissa.it}\\
{\it INFN, Sezione di Trieste and Scuola Internazionale}\\
{\it Superiore di Studi Avanzati (SISSA)} \\
{\it Via Beirut 4, I-34013 Trieste, Italy}
\\
\\
Chuan-Jie Zhu\thanks{E-mail: zhucj@itp.ac.cn} \\
{\it Institute of Theoretical Physics, Chinese Academy of}\\
{\it Sciences, P. O. Box 2735, Beijing 100080, P. R. China}}

\maketitle

\begin{abstract}
In this note we derive an explicit modular invariant
formula for the two loop 4-point amplitude in superstring
theory in terms of a multiple integral (7 complex integration variables)
over the complex plane which is shown to be convergent.
We consider in particular the case 
of the leading term for vanishing momenta of the four graviton amplitude,
which would correspond to the two-loop correction of the $R^4$ term
in the effective Action. The resulting expression is not positive definite
and could be zero, although we cannot see that it vanishes.   

\end{abstract}

\newpage

\section{Introduction and Summary}
Inspired by a  recent argument \cite{Green,GreenSethi} about the $R^4$ term in
superstring
theory, it is necessary to reconsider the finiteness of the superstring
two-loop 4-graviton amplitude \cite{IengoZhu}, especially for the case of
vanishing 
momenta which was not discussed in detail previously in \cite{IengoZhu}.
The arguments
given in \cite{IengoZhu} were not complete. We can extend these
arguments
and make them precise \cite{Zhu}. However we will take a different route here.

First we derive an explicit modular invariant formula for the two-loop
4-point amplitude.
We remind that the genus-2 Riemann surface, which is the appropriate world
sheet for two loops, can be described in full generality by means of the
hyperelliptic formalism. This is based on a representation of
the surface as two sheet covering of the complex plane described by the
equation:
\begin{equation}
y^2(z) = \prod_{i=1}^6 ( z- a_i), \label{covering}
\end{equation}
The complex numbers $a_{i}$,
$(i=1,\cdots,6)$ are the six branch points, by going around them one passes 
from one sheet to the other. 
Three of them represent the moduli of the genus 2 Riemann
surface over which the integration is performed, while the other three can be
arbitrarily fixed. In fact, the two-loop formula, which we obtained by
implementing in the hyperelliptic formalism the general algorithm 
\cite{verlinde} for a multi-genus worldsheet,
 is expressed as an
integral over the Riemann surface moduli, i.e. three branch points, and over
the four (complex) position
of the vertex operators. The whole integrand, including the measure, is
$SL(2,C)$ invariant and thus three points (i.e. the remaining branch points)
are arbitrarily fixed.   
 Modular invariance in this language amounts to invariance under
permutations of
$a_{i}$, $(i=1,\cdots,6)$, and it is not explicitely apparent in
the expression for the amplitude given in \cite{IengoZhu}.
One of the main goal of this paper is to get it manifest. 

Our strategy is the following. 
In our previous calculation 
we fixed 3 branch points by using the $SL(2,C)$  
invariance of the hyperelliptic representation of genus 2 Riemann surface.
Here instead (see Sect. 2) we choose to
fix 3 out of the 4 vertex operator points and integrate all the 6 branch
points. We do it by making a suitable $SL(2,C)$ change of variables on
our previous formula. 
Amazingly
an explicit modular invariant formula is obtained for the 4-point
amplitude. This formula
is  like the Koba-Nielsen formula for the tree amplitude in string theory, 
in that it is in fact represented as an integral over the whole complex
plane  of one of the four vertex positions (the other three being fixed).
Furthermore there is also the integration over the whole complex plane
of the six branch points which appear symmetrically in the integrand.   

With this new formula in hand we study its finiteness property
by making a convenient choice
of the two arbitrary parameters from supercurrent insertion points 
(some basic facts about this ingredient of the construction of the
amplitude are recalled in Sect. 2.1).  We get thus the
following formula for the zero momentum limit of the four graviton
amplitude in Type II (A or B) Superstring theory
($z_{1,2,3}=z^0_{1,2,3}$ arbitrarily fixed):
\begin{eqnarray}
 AII_0 = c_{II}\, K\, V \int
{ \prod_{i=1}^6 d^2 a_i \over
{T}^5 \prod_{i<j}^6 |a_{ij} |^2 }\,
 {d^2 z_4  \over | y(z_4)|^2  }
\, { (z_1^0- z_4) (\bar{z}_2^0 - \bar{z}_4) \over
|y(z_1^0)y(z_2^0)y(z_3^0)|^2 }
\qquad \qquad & &
\nonumber \\
 \quad  \times
\left( { 1\over 2} \sum_{i=1}^6 { 1\over z_1^0 - a_i } - \sum_{l\neq 1 }^4
{1\over z_1^0 - z_l} \right)
\left(  { 1\over 2} \sum_{i=1}^6 { 1\over \bar{z}_2^0 - \bar{a}_i } -
\sum_{l\neq 2 }^4
{1\over \bar{z}_2^0 - \bar{z}_l} \right) , & &
\label{Eqlastzzz}
\end{eqnarray}
where $T$ (the determinant of the ``period matrix" of the surface) is a
function of $a_i,\bar a_i$ given in eq. (\ref{Eqtwo}) below, 
$K$ is the standard 4-graviton kinematical factor \cite{GreenSchwarz,IengoZhu} 
(see Appendix B for details and its relation with the $R^4$ term)
and  $V$ is a constant factor:
\begin{equation}
V = - { 1\over 16} |z_1^0 - z_2^0|^4 (z_1^0 -z_3^0)^2 (\bar{z}_1^0
-\bar{z}_3^0)
(z_2^0 -z_3^0)(\bar{z}_2^0 -\bar{z}_3^0)^2.
\end{equation}

The constant $c_{II}$ in front of the amplitude in eq. (\ref{Eqlastzzz})
can be determined by the 
unitarity relations which for generic momenta relate the two-loop 
expression with the one-loop and the tree level amplitudes, and thus it is
nonzero.
 
One can check that the integrand of the above formula is $SL(2,C)$ invariant
(including the measure), i.e. invariant under a simultaneous transformation 
of all the $a_i$'s and $z_l$'s of the kind $w \to (aw+b)/(cw+d)$ with
$ad-bc=1$. Every factor transforms nontrivially and it is
rather amazing to see how the whole formula is calibrated. 
It is also manifestly invariant under permutations
of the six branch points $a_{1,\cdots ,6}$, i.e. it is manifestly modular
invariant. In Sect. 3 we make a thorough study of all the possible
boundaries of the moduli space and prove that the amplitude is finite,
i.e. it is expressed as a convergent 7-(complex)-dimensional integral.

It is not clear whether the result of the integration is zero as it should
be for in agreement with the arguments given in refs. \cite{Green, GreenSethi} 
about the absence of $n$-loop
(with $n$ larger than 1) contributions to the $R^4$ term.
The above expression for the integrand is certainly not positive definite
(we will see in Sect. 2.2 that attemping to get a positive definite
integrand by adding some total derivative, as it was uncorrectly argued in 
\cite{IengoZhu}, is not possible due to boundary terms). 
Still it does not appear to be zero in an obvious way. This can be
contrasted with the fact that using the same formalism we have verified
other nonrenormalization theorems \cite{IengoZhu-previous},  namely
the vanishing of the amplitude with less than 4 vertices. In those cases 
the integrand itself was found to be zero.

\section{A Modular Invariant Form of the Amplitude}

\subsection{Heterotic String Amplitude}
To simplify the presentation we will first discuss the amplitude for heterotic
string.  The two loop 4-point amplitude in heterotic string theory 
obtained in \cite{IengoZhu, ZhuDr} is 
\begin{equation}
AH_0  =  c_H\, K_H \,  \int { d^2 a_1 d^2 a_2 d^2 a_3 \,
|a_{45}\,a_{46}\,a_{56}|^2 \over 
T^5 \prod_{i<j}^6 |a_{ij}|^2 } 
\prod_{l=1}^4 {d^2 z_l (r-z_l)\over  y(z_l) } 
 I(r) \bar{F}(\bar{a}, \bar{z}),
\label{Eqahzero}
\end{equation}
where $K_H$ is a kinematic factor depending on the left sector particle contents
and $\bar{F}(\bar{a}, \bar{z})$ is a function
denoting the contribution from left sector of the heterotic string theory.
The complex numbers $a_{i}$,
$(i=1,\cdots,6)$ and the function $y(z)$ which describe the Riemann
surface have been introduced in the first Section, see eq. (\ref{covering}). 
Three of the six branch points, say $a_{1,2,3}$, represent the moduli of 
the genus 2 Riemann surface
over which the integration is performed, 
while $a_{4,5,6}$ are arbitrarily fixed.
Modular invariance in this language amounts to invariance under permutations of 
$a_{i}$, $(i=1,\cdots,6)$, and it is 
not explicitely apparent in eq. (\ref{Eqahzero}).
One of the main goal of this paper is to get it manifest. Let us remind that 
eq. (\ref{Eqahzero}) 
is obtained after summing over the even spin structures with signs uniquely fixed
by the requirement of modular invariance \cite{GavaISotkov}.
There is no contribution from the odd spin structure. In fact, the four
vertices and the two supercurrent insertion (which we recall below)
could just provide the required ten fermionic zero modes, resulting in a
contribution proportional to the completely antisymmetric ten dimensional
tensor. Since it will be contracted with the four momenta, which are
linearly dependent, the result will be identically zero.    

We also recall
that  $ \bar{F}(\bar{a}, \bar{z})$ is symmetric under permutations of
$\bar a_{i}$, $(i=1,\cdots,6)$.   
The other functions appearing in (\ref{Eqahzero}) are
\begin{equation}
T (a_i, \bar{a}_i)  =  \int { d^2 z_1 d^2 z_2
 |z_1 - z_2|^2 \over |y(z_1)y(z_2)|^2 }, 
\label{Eqtwo}
\end{equation}
which is proportional to
the determinant of the period matrix (see ref. \cite{ZhuDr}), and 
\begin{eqnarray}
I(r)& = & -{1\over 2} \sum_{i<j}^6 { 1\over r -a_i}
 {1\over r -a_j} - 
{1\over 4} \sum_{i<j}^3 { 1\over r -a_i} {1\over r -a_j} 
\nonumber \\
& & + {1\over 8} \left( \sum_{i=1}^6 {1 \over r - a_i} - 2 
 \sum_{i=1}^3 {1 \over r - a_i} 
\right) \sum_{l=1}^4 { 1\over r -z_l}
\nonumber \\
& & + {1\over 4}\sum_{i=1}^6 {1\over r -a_i} \sum_{i=1}^3 
{1\over r -a_i} - {5\over 4} 
\sum_{i=1}^6 {1\over r -a_i} 
{\partial \over \partial a_i} \ln T, 
\label{Eqthree}
\end{eqnarray}

As we proved in \cite{IengoZhu, ZhuDr} the amplitude $AH_0$ is
independent of the
parameter $r$. This parameter $r$ represents the arbitrary point of insertion of
the two supercurrents (also called picture-changing operators) 
as appropriate for genus 2 where there are two zero modes of the $\beta$
superghost.  We have chosen to insert those two supercurrents 
at the same point $r$ in
the upper and lower sheet of the double covering of the complex plane 
described by
eq. (\ref{covering}). The independence of the result from $r$ is in 
agreement with 
a general derivation in ref. \cite{verlinde} where it is shown in general 
that a displacement of $r$ amounts to an irrelevant total derivative in the
moduli space.  

In fact, we note that $I(r)$ is a rational function of $r$. 
By a standard argument from complex analysis or by explicit computation we
have
\begin{equation}
I(r)\, \prod_{l=1}^4 (r - z_l)= \sum_{i=1}^6 I_i(r) \prod_{l=1}^4
(a_i -z_l) +I_{\infty}, 
\label{Eqfoura}
\end{equation}
where $I_{\infty}$ is independent of $r$ (see eq. (\ref{Eqfive}) in Appendix A)
and $I_i(r)$ contains the pole singularities for $r \to  a_i$.

As we proved in \cite{IengoZhu}, all these singular terms give  total
derivatives 
when we insert eq. (\ref{Eqfoura}) into the 4-point amplitude (\ref{Eqahzero}).
For instance the formula relevant for $I_i(r)$, $(i=1,2,3)$ is
(see the Appendix A for further details):
\begin{equation}
 I_i(r)\prod_{l=1}^4 {(a_i-z_l)\over y(z_l) }
{1\over T^5 \prod_{k<l}^6 
a_{kl} } = 
{1\over 4} {\partial \over \partial a_i} 
\left\{ { 1\over r - a_i} {1\over  T^5 \prod_{k<l}^6 
a_{kl}} \prod_{l=1}^4 {(a_i-z_l) \over y(z_l)}
\right\} ,
\label{Eqnine}
\end{equation}

Similar formulas hold also for $I_i(r)$, $(i=4,5,6)$ (eq. (\ref{Eqtena}) in Appendix A). 
These total derivative terms give zero contribution when integrated over the
moduli as in eq. (\ref{Eqahzero}), since we proved in \cite{IengoZhu, ZhuDr}
that there are no possible boundary contributions. Therefore 
$AH_0$ is independent from $r$. 

Next, we note that the amplitude (\ref{Eqahzero}) is $SL(2,C)$ invariant, 
i.e. a simultaneous $SL(2,C)$ transformation
on all the integration variables $a_{1,2,3}$, $z_{1,2,3,4}$ and the fixed
point $a_{4,5,6}$ and $r$ leaves $AH_0$ invariant, i.e.
\begin{eqnarray}
AH_0  & = & c_H\,K_H\, \int { d^2 {S}(a_1) d^2 {S}(a_2) d^2
{S}(a_3) \,
\prod_{i<j=4}^6 |{S}(a_i) - {S}(a_j)|^2 \over 
{T}^5 \prod_{i<j}^6 |{S}(a_i) - {S}(a_j)|^2 } 
\nonumber \\
& & \times 
\prod_{l=1}^4 {d^2 {S}(z_l) ({S}(r)-{S}(z_l))\over  y({S}(z_l)) } 
 I({S}(r)) \bar{F}(\bar{S}(\bar{a}),\bar{S}(\bar{z}) ),
\label{Eqahzeroo}
\end{eqnarray}
where $S$ denotes an $SL(2,C)$ transformation. Here the functions
$T$, $y({S}(z))$ and $I(S(r))$ have in their definitions with
$a_i$ changed to ${S}(a_i)$.
$S(x)$ depends on three parameters $q_{1,2,3}$ and thus we set
\begin{equation}
S(x)\equiv S_q(x) = { (q_1 -q_2 )(x -q_3) \over (q_1-q_3)(x -q_2)}.
\end{equation}
Now we insert the following identity
\begin{equation}
\int  \prod_{i=1}^3 d^2 q_i \, \prod_{i=1}^3 \delta^2( S_q(z_i) - Z_i^0) \, 
{\prod_{i<j=1}^3 |Z_i^0-Z_j^0|^2 \over \prod_{i<j=1}^3 |q_i^0-q_j^0|^2 } =1,
\end{equation}
into  eq. (\ref{Eqahzeroo}) and we get
\begin{eqnarray}
AH_0 & = & 
c_H\,K_H\,  \int { d^2 {S}(a_1) d^2 {S}(a_2) d^2 {S}(a_3) \,
\prod_{i<j=4}^6 |{S}(a_i) - {S}(a_j)|^2 \over 
{T}^5 \prod_{i<j}^6 |{S}(a_i) - {S}(a_j)|^2 } 
\nonumber \\
& & \times 
\prod_{l=1}^4 {d^2 {S}(z_l) ({S}(r)-{S}(z_l))\over  y(
{S}(z_l)) }  
 \prod_{i=1}^3 d^2 q_i \, \prod_{i=1}^3 \delta^2( S_q(z_i) - Z_i^0) 
\nonumber \\
& & \times  
{\prod_{i<j=1}^3 |Z_i^0-Z_j^0|^2 \over \prod_{i<j=1}^3 |q_i^0-q_j^0|^2 } 
 I({S}(r)) \bar{F}(\bar{{S}}(\bar{a}),
\bar{{S}}(\bar{z}) ).
\end{eqnarray}
Now  taking ${S} = S_q$ and noticing the following relation
\begin{equation}
\prod_{i=1}^4 d q_i \, { \prod_{i<j=4}^6 (S_q(a_i) - S_q(a_j) ) \over 
\prod_{i<j=1}^3 (q_i-q_j) } = \prod_{i=4}^6 d S_q(a_i) ,
\end{equation}
we get 
\begin{eqnarray}
AH_0 & = & c_H\,K_H\,  \int 
{ \prod_{i=1}^6 d^2 A_i \over 
{T}^5 \prod_{i<j}^6 |A_i - A_j |^2 } 
\prod_{l=1}^4 {d^2 Z_l({S}_A(r)-Z_l)\over  Y(Z_l) }  
\nonumber \\
& & \times  
\prod_{i=1}^3\delta^2( Z_i- Z_i^0) 
{\prod_{i<j=1}^3 |Z_i^0-Z_j^0|^2 }
 I({S}_A(r)) \bar{F}(\bar{A},  \bar{z}) ),
\label{Eqlasta}
\end{eqnarray}
where we have renamed $S_q(a_i)$ to be $A_i$, $S_q(z_l)$ to be 
$Z_l$ and $Y^2(Z) = \prod_{i=1}^6(Z-A_i)$.
Also we changed the name of $SL(2,C)$ transformation of
$S_q$ to be $S_A$, since $q_{1,2,3}$ are fixed in terms of $A_{4,5,6}$. 
Explicitly we have 
\begin{equation}
S_q(r)=S_A(r) = {
A_6(A_4-A_5) r - A_4(A_6-A_5) \over (A_4-A_5) r - (A_6-A_5) } .
\end{equation}

From eq. (\ref{Eqlasta}) we see that because of the 3 $\delta$ functions 
the integration variables $Z_{1,2,3}$ are fixed and  all the $A_{1,\cdots,6}$
are integrated.  Nevertheless this formula is not explicitly modular
invariant, 
i.e. it's not invariant under any interchange of $A_i$ and $A_j$. In what
follows we
will put it into a modular invariant form.

Setting 
\begin{eqnarray}
I_M(x) & = & 
{1\over 4} \sum_{i=1}^6 { 1\over (x- A_i)^2 }-
{1\over 4} \sum_{i<j}^6 { 1\over x -A_i}
 {1\over x -A_j} -   {1\over 8} \sum_{i=1}^6 {1 \over x- A_i}  
 \sum_{l=1}^4 { 1\over x -Z_l}
\nonumber \\
& & + {1\over 4}\sum_{k<l=1}^4 {1\over x -Z_k} {1\over x -Z_l} - {5\over 4} 
\sum_{i=1}^6 {1\over x-A_i} 
{\partial \over \partial A_i} \ln T, 
\label{Eqthreem}
\end{eqnarray}
we have the following identity
\begin{eqnarray}
& & I(S_A(r)) \prod_{k=1}^4  { S_A(r) - Z_k\over Y(Z_k) } \, 
{ 1\over T^5 \prod_{i<j=1}^6 A_{ij} }  = 
I_M(x) \prod_{k=1}^4  { x - Z_k\over Y(Z_k) } \, 
{ 1\over T^5 \prod_{i<j=1}^6 A_{ij} }  
\nonumber \\
& &  - { 1\over 4} \sum_{i=1}^6{\partial \over \partial A_i } 
\left[ \left( { 1\over A_i - S_A(r) } - {1\over  A_i - x} \right)
 \, \prod_{l=1}^4  
{ A_i - Z_l\over Y(Z_l) } \, { 1\over T^5 \prod_{j<k=1}^6 A_{jk} } \right].
\label{Eqthreemm}
\end{eqnarray}
By using this identity in eq. (\ref{Eqlasta})
and by dropping all the total derivatives terms, we finally get 
\begin{eqnarray}
AH_0 & = & c_H\,K_H\, \int 
{ \prod_{i=1}^6 d^2 A_i \over 
{T}^5 \prod_{i<j}^6 |A_i - A_j |^2 } 
\prod_{l=1}^4 {d^2 Z_l (x - Z_l) \over  Y(Z_l) }  
\nonumber \\
& & \times  
\prod_{i=1}^3\delta^2( Z_i- Z_i^0) 
{\prod_{i<j=1}^3 |Z_i^0-Z_j^0|^2 } \, I_M(x)
 \bar{F}(\bar{A},  \bar{z}) ).
\label{Eqlastd}
\end{eqnarray}
Notice that $I_M(x)$ is a modular invariant function, i.e. invariant under
any permutation of the branch points $A_i$, $(i=1,\cdots,6)$. So we obtain 
a modular invariant formula for the two loop 4-particle amplitude
in heterotic string theory. It is also independent of the parameter $x$ as
one can easily check that changing $x$ amounts to irrelevant total derivatives,
as we have done in Appendix A.

\subsection{Type II Superstring Amplitude}
For type II superstring theory, we can use the same method to get a modular
invariant
formula. Let us start with the two loop 4-point amplitude (with vanishing
momenta)
in type II superstring theory
\cite{IengoZhu, ZhuDr}:
\begin{eqnarray}
AII_0  &=&  c_{II}\, K \int { d^2 a_1 d^2 a_2 d^2 a_3 \,
|a_{45}\,a_{46}\,a_{56}|^2 \over 
T^5 \prod_{i<j}^6 |a_{ij}|^2 } 
\prod_{l=1}^4 {d^2 z_l (r-z_l)(\bar{s}-\bar{z}_l)
\over
|y(z_l)|^2 } 
\nonumber \\
& & \times \left\{ I(r)\bar{I}(\bar{s}) + 
{5\over 4}\left( {\pi \over T \, y(r)\bar{y}(\bar{s})
} \int {d^2 w (r-w)
(\bar{s}-\bar{w})\over |y(w)|^2 }\right)^2 \right\},
\label{Eqone}
\end{eqnarray}
where $K$ is the standard 4-graviton kinematical factor
\cite{GreenSchwarz,IengoZhu}.
The complete amplitude and $K$ factor is given in Appendix B.
Now there are two
arbitrary parameters from supercurrent-insertion points,
$r$ from the right sector and $\bar s$ from the left sector. Again, the resulting
amplitude $AII_0$ is independent of them.

Now we perform the same trick of performing a $SL(2,C)$ transformation on (\ref{Eqone})
as in the heterotic string case and
obtain the following:
\begin{eqnarray}
& & AII_0  =  c_{II}\,K \,\int 
{ \prod_{i=1}^6 d^2 A_i \over 
{T}^5 \prod_{i<j}^6 |A_{ij} |^2 } 
\prod_{l=1}^4 {d^2 Z_l({S}_A(r)-Z_l)(\bar{S}_A(\bar{s}) - \bar{Z}_l) 
\over  |Y(Z_l)|^2 }  
\nonumber \\
& &\qquad  \times  
\prod_{i=1}^3\delta^2( Z_i- Z_i^0) 
{\prod_{i<j=1}^3 |Z_i^0-Z_j^0|^2 } \Big\{
 I({S}_A(r)) \bar{I}(\bar{S}_A(\bar{s})  
\nonumber \\
& & + \left. {5\over 4} \left(
{\pi \over T \, Y(S_A(r))\bar{Y}(\bar{S}_A(\bar{s}))
} \int {d^2 v (S_A(r)-v)
(\bar{S}_A(\bar{s})) -\bar{v})\over |Y(v)|^2 }\right)^2 \right\}.
\label{Eqaddc}
\end{eqnarray}

To proceed further we need some formulas reported in the Appendix B.

By using eq. (\ref{finalAppendix}) of the Appendix B in (\ref{Eqaddc}), it is
now easy to derive an explicit 
modular invariant formula for type II superstring by dropping all total
derivative terms.
We get  
\begin{eqnarray}
& & AII_0   = c_{II}\,   K\, \int 
{ \prod_{i=1}^6 d^2 A_i \over 
{T}^5 \prod_{i<j}^6 |A_{ij} |^2 } 
\prod_{l=1}^4 {d^2 Z_l(x- Z_l) (\bar{w} - \bar{Z}_l) \over | Y(Z_l)|^2  }  
\nonumber \\
& & \quad \times \prod_{i=1}^3\delta^2( Z_i- Z_i^0) 
{\prod_{i<j=1}^3 |Z_i^0-Z_j^0|^2 } 
\nonumber \\
& &\quad  \times  \left\{ I_M(x) \bar{I}_M(\bar{w})+ {5 \over 4} 
\left( {\pi \over T \, Y(x)\bar{Y}(\bar{w})
} \int {d^2 v (x-v)
(\bar{w}-\bar{v})\over |Y(v)|^2 }\right)^2 \right\}.
\label{Eqlastz}
\end{eqnarray}
where $I_M(x)$ is the same as in eq. (\ref{Eqthreem}). Again, $x$ and $\bar w$
are arbitrary and the result does not depend on them.

We note  that we can't set $x=w$ in eq. (\ref{Eqlastz}) because with
$x=w$ in eq. (\ref{Eqlastz})  $AII_0$ is divergent. The important point
to note
here is the following: the total derivative terms (see eq. (\ref{finalAppendix}))
which would be dropped are actually
in this case non-vanishing
and divergent. This can be clearly demonstrated by the following 
computation ($A_i-x = r \, e^{i\theta}$):
\begin{eqnarray}
\int_{\epsilon<|A_i-x|<\Lambda}  d^2 A_i {\partial \over  \partial  A_i}\left[
{ 1\over A_i-x}  {f(A,\bar{A})} \bar{I}_M(\bar{w}) \right]  \hskip 4cm & & 
\nonumber \\
  = \int_{\epsilon}^{\Lambda} r\, d\, r \int_0^{2 \pi}d\theta 
\, e^{-i\theta}\, \left( {\partial \over \partial r} - {i\over r}\,
{\partial\over \partial 
\theta} \right) \left[{ 1\over r\, e^{i\theta}}  {f(A,\bar{A})} 
\bar{I}_M(\bar{w}) \right]  \qquad & &
\nonumber \\
 = \int_{\epsilon}^{\Lambda}  d\, r \int_0^{2 \pi} d\theta \left( 
e^{-2 i \theta } {\partial\over \partial r }  {f(A,\bar{A})}
\bar{I}_M(\bar{w}) + 
{1\over r}\, {\partial\over \partial \theta } \left[ e^{- 2 i \theta}\,
{f(A,\bar{A})} 
\bar{I}_M(\bar{w}) \right] \right).& &
\end{eqnarray}
In the above equation the last term is a total derivative in $\theta$ and
can be dropped.
On the other hand the first term is a total derivative in $r$ and its
integration
over $r$ gives two terms evaluated at the points $r=\epsilon$ and $r=\Lambda$:
\begin{equation}
\int_{0}^{2 \pi} d \theta \left( e^{- 2 i \theta}\,  {f(A,\bar{A})} 
\bar{I}_M(\bar{w}) |_{r = \Lambda} - 
e^{- 2 i \theta}\,  {f(A,\bar{A})} \bar{I}_M(\bar{w}) |_{r = \epsilon}
\right) .
\end{equation}
Here we only need to consider the second term because the first term is
just an 
artifact of cutoff.  When $w\neq x$,  the second term is regular for
$r=\epsilon\to 0$
and so there is no boundary term. When $w =x$, the term ${1\over (\bar{A}_i-
\bar{w})^2}$ from $\bar{I}_M(\bar{w})$ gives a contribution
which is ${2 \pi \over \epsilon^2}$
after integration over $\theta$ which is divergent for $\epsilon \to 0 $
while the other terms 
give non-vanishing but finite contributions. 

\section{The finiteness of the amplitude}

Now we study the finiteness property of the amplitude $AII_0$ in
(\ref{Eqlastz}). First
we make the following choice of the two arbitrary parameters $x$ and $w$
(which have 
their origin as the insertion point of the supercurrents): $x= Z_1^0 $ and
$w = Z_2^0$.
With this choice the amplitude $AII_0$ simplifies greatly:
\begin{eqnarray}
 AII_0 =  c_{II}\, K\, V \int 
{ \prod_{i=1}^6 d^2 A_i \over 
{T}^5 \prod_{i<j}^6 |A_{ij} |^2 }\, 
 {d^2 Z_4  \over | Y(Z_4)|^2  }  
\, { (Z_1^0- Z_4) (\bar{Z}_2^0 - \bar{Z}_4) \over
|Y(Z_1^0)Y(Z_2^0)Y(Z_3^0)|^2 }
\qquad \qquad & &
\nonumber \\
 \quad  \times 
\left( { 1\over 2} \sum_{i=1}^6 { 1\over Z_1^0 - A_i } - \sum_{l\neq 1 }^4 
{1\over Z_1^0 - Z_l} \right)
\left(  { 1\over 2} \sum_{i=1}^6 { 1\over \bar{Z}_2^0 - \bar{A}_i } -
\sum_{l\neq 2 }^4 
{1\over \bar{Z}_2^0 - \bar{Z}_l} \right) , & &
\label{Eqlastzz}
\end{eqnarray}
where $V$ is a constant factor:
\begin{equation}
V = - { 1\over 16} |Z_1^0 - Z_2^0|^4 (Z_1^0 -Z_3^0)^2 (\bar{Z}_1^0
-\bar{Z}_3^0) 
(Z_2^0 -Z_3^0)(\bar{Z}_2^0 -\bar{Z}_3^0)^2.
\end{equation}

To prepare the discussion of finiteness of the above amplitude 
we first note that due to the angular integration,
the factor $(Z_1^0-Z_4)(\bar{Z}_2^0 -\bar{Z}_4)$ times the second line of
eq. (\ref{Eqlastzz}) can be omitted. Although this factor may be singular
for $A_i \to 
Z_{1,2}^0$ (it is not singular for $Z_4 \to Z_{1,2}^0$),
the singular terms only appear either in the holomorphic part or the
anti-holomorphic part.  The integration over the angular variable then
gives  a contribution
which is always not more singular than the contribution without this
factor. So for the
purpose  of discussing finiteness we can just consider the finiteness of
the truncated  
expression:
\begin{equation}
 AII_T =   \int 
{ \prod_{i=1}^6 d^2 A_i \over 
{T}^5 \prod_{i<j}^6 |A_{ij} |^2 }\, 
 {d^2 Z_4  \over | Y(Z_4)|^2  }  
\, {1 \over |Y(Z_1^0)Y(Z_2^0)Y(Z_3^0)|^2 }.
\label{Eqzxc}
\end{equation}

To discuss the finiteness property of the amplitude $AII_T$ in (\ref{Eqzxc}) 
we keep the fixed points $Z_i^0 $ at generic points and classify all the
boundaries 
into various cases. 

In the first case a) all the branch points are kept away
from the fixed
points $Z_i^0$ and we  use
the following parametrization:
\begin{eqnarray}
A_1 & = & A_1,
\\
A_2 & = & A_1 + u ,
\\
A_i & = & A_1  + u \,v_{i-2}, \qquad i = 3, \cdots, n \leq 6,
\end{eqnarray}
and consider the limit $u \to 0$ and $v_i$'s finite. The integration over
all $A_i$ ($i=1,\cdots,6$) changes to an integration over $u$, $A_1$, 
$v_i$ and the rest $A_j$'s.

In the second case b) a subset of the branch
points approach one of the points of $Z_i^0$ and we use the following
parametrization:
\begin{eqnarray}
A_1 & = & Z_i^0 + u ,
\\
A_j & = & Z_i^0  + u \, v_{j-1}, \qquad j = 2, \cdots, n \leq 6,
\end{eqnarray}
and also consider the limit $u \to 0$ and $v_i$'s finite. The integration over
all $A_i$ ($i=1,\cdots,6$) changes to an integration over $u$, 
$v_i$ and the rest $A_j$'s. 

The third case is a combination of
the first 
two cases and its finiteness is proved if we can prove the finiteness in
the first two
cases. In each cases the integration over $Z_4$ must also be considered to
obtain
the most singular behaviour.

It is straightforward to compute the behaviour of the various factors under
various
degeneration limits. In particular the function $T$ under various
degeneration limit
$u \to  0 $ behaves as follows:
\begin{equation} 
T \sim  \left\{ \begin{array}{ll}
 \ln |u| ,  & A_2 \to A_1 , \\
 {1\over |u|},  & A_{2,3} \to A_1,\\
 {\ln |u| \over |u|^2 }, & A_{2,3,4} \to A_1,\\
 {1 \over |u|^4 }, & A_{2,\cdots,5} \to A_1,\\
 {1 \over |u|^6 }, & A_{2,\cdots,6} \to A_1. 
\end{array}
\right. 
\end{equation}

In the first case the factor ${1\over |Y(Z_1^0)Y(Z_2^0)Y(Z_3^0)|^2}$ is
non-singular.
The behaviour of the other factors is given in the following table:

\vskip 1cm
\begin{tabular}{|c|c|c|c|c|c|}
 \hline
 & ${1\over T^5}$ & $\prod_{i=1}^6d^2 A_i$ & ${1\over \prod_{i<j}^6
|A_{ij}|^2}$ 
& $\int {d^2 Z_4 \over |Y(Z_4)|^2}$ & All \\
\hline
$A_2 \to A_1 $ & ${1\over (\ln |u|)^5}$ & $d^2 u$ & ${1\over |u|^2}$ 
& $\ln |u|$ & 
${ d^2 u \over |u|^2}\, {{1\over (\ln |u|)^4}}$ \\
\hline
$A_{2,3}\to A_1$ & $|u|^5$ & $|u|^2  d^2 u$ & ${1\over |u|^6 }$ & 
${1\over |u|}$ & 
$d^2 u $\\
\hline 
$A_{2,3,4}\to A_1$ & $|u|^{10} \over (\ln|u|)^5$ & $|u|^4  d^2 u$ &
${1\over |u|^{12}
 }$ &  ${1\over |u|^2}$ & $d^2 u \, {{1\over (\ln |u|)^5}} $\\
\hline 
$A_{2,\cdots,5}\to A_1$ & $|u|^{20} $ & $|u|^6  d^2 u$ & ${1\over |u|^{20}
 }$ &  ${1\over |u|^3}$ & $ |u|^3 d^2 u $\\
\hline 
$A_{2,\cdots,6}\to A_1$ & $|u|^{30} $ & $|u|^8  d^2 u$ & ${1\over |u|^{30}
 }$ &  ${1\over |u|^4}$ & $|u|^4 d^2 u $\\
\hline
\end{tabular}
\vskip 1cm  
In the last column we have put all factors together and  one sees quite
clearly that
the integration over $u$ gives a finite contribution from the corner $u
\sim 0$.
This proves that the amplitude $AII_0$ is finite in the first case a)
degeneration limit.

In the second case b) only one $Y(Z_i^0)$ in the factor ${1\over |Y(Z_1^0)Y(Z_2^0)
Y(Z_3^0)|^2}$ gives singular contribution. This contribution and the
behaviour of the
other factors is given in the following table:

\vskip 1cm
\noindent
\begin{tabular}{|c|c|c|c|c|c|c|}
 \hline
 & ${1\over T^5}$ & $\prod_{i=1}^6d^2 A_i$ & ${1\over \prod_{i<j}^6
|A_{ij}|^2}$ 
& $\int {d^2 Z_4 \over |Y(Z_4)|^2}$ & ${1\over |Y(Z_i^0)|^2}$ & All \\
\hline
$A_{1} \to Z_i^0$ & ${1}$ & $d^2 u$ & ${1}$  & 1 
& ${1\over |u|}$ & ${d^2 u \over |u|}$ \\
\hline
$A_{1,2} \to Z_i^0$ & ${1\over (\ln |u|)^5}$ & $|u|^2 d^2 u$ & ${1\over
|u|^2}$ 
& $\ln |u|$ & ${1\over |u|^2}$ & 
${ d^2 u / |u|^2\over (\ln |u|)^4}$ \\
\hline
$A_{1,2,3}\to Z_i^0$ & $|u|^5$ & $|u|^4  d^2 u$ & ${1\over |u|^6 }$ & 
${1\over |u|}$ & ${1\over |u|^3}$ & $d^2 u\over |u| $\\
\hline 
$A_{1,\cdots,4}\to Z_i^0$ & $|u|^{10} \over (\ln|u|)^5$ & $|u|^6  d^2 u$ &
${1\over |u|^{12}
 }$ & ${1\over |u|^2}$ &  ${1\over |u|^4}$ & ${ {d^2 u / |u|^2\over (\ln
|u|)^5}}$  \\
\hline 
$A_{1,\dots,5}\to Z_i^0$ & $|u|^{20} $ & $|u|^8  d^2 u$ & ${1\over |u|^{20}
 }$ &  ${1\over |u|^3}$ &${1\over |u|^5}$ &  $  d^2 u $\\
\hline 
$A_{1,\cdots,6}\to Z_i^0$ & $|u|^{30} $ & $|u|^{10}  d^2 u$ & ${1\over
|u|^{30}
 }$ &  ${1\over |u|^4}$ & ${1\over |u|^6}$ & $ d^2 u $\\
\hline
\end{tabular}
\vskip 1cm
In the last column we have put all factors together and  one sees quite
clearly that
the integration over $u$ gives a finite contribution from the  corner $u
\sim 0$.
This proves that the amplitude $AII_0$ is also
finite in the second case b) degeneration limit. This completes our proof that the
modular invariant two-loop 4-particle superstring amplitude is finite.

\section *{Acknowledgments}

C.-J. Zhu is supported in part by funds from National Science Foundation of
China
and Pandeng Project. 

\section *{Appendix A}

In this appendix we gave some details \cite{IengoZhu, ZhuDr}
in proving the independence of the
heterotic  string amplitude on the parameter $r$.  Let us first recall eq. (\ref{Eqfoura})
here:
\begin{equation}
I(r)\, \prod_{l=1}^4 (r - z_l)= \sum_{i=1}^6 I_i(r) \prod_{l=1}^4
(a_i -z_l) +I_{\infty},
\label{Eqfour}
\end{equation}
where $I_{\infty}$ is given as follows \cite{IengoZhu}:
\begin{eqnarray}
I_{\infty} & = & - {1\over 2} \sum_{i<j}^6 a_i a_j - {1\over 4}
\sum_{i<j}^6 a_i a_j+ {1\over 8} (\sum_{i=1}^6 a_i -2\, \sum_{i=1}^3
a_i)\sum_{l=1}^4 z_l
 \nonumber \\
& &  + {1\over 4} \sum_{i=1}^6 a_i \sum_{j=1}^3 a_j
-{5\over 4} \sum_{i=1}^6 \Big( a_i^2 + a_i^3 { \partial\over \partial a_i}
\ln T\Big),   
\label{Eqfive}
\end{eqnarray}
and $I_i(r)$ are defined as the singular terms of $I(r)$ times some
factors as $r \to a_i$:
\begin{equation}
I(r)\, \prod_{l=1}^4 (r - z_l) \to I_i(r) \, \prod_{l=1}^4 (a_i -z_l),
\qquad \hbox{for $r \to a_i$.}
\label{Eqsix}
\end{equation}
Explicitly we have:
\begin{eqnarray}
I_i(r) & = & {1\over 4} {1\over (r -a_i)^2} - {1\over 4}\,
{1\over r - a_i} \sum_{j\neq i}{ 1\over a_i-a_j}
\nonumber \\
& & +
{1\over 8} {1\over r -a_i}\, \sum_{l=1}^4 { 1\over a_i - z_l} - { 5 \over 4}
{1\over r -a_i}  {\partial \over \partial a_i} \ln T, \qquad
i=1,2,3,
\label{Eqseven}
\\
I_4(r) & = & {1\over r -a_4} \left(
-{1\over 2}\left( {1\over a_4-a_5} + {1\over a_4 -a_6} \right) \right.
\nonumber \\
& & \left. -{1\over 4}
\sum_{i=1}^3 {1\over a_4 -a_i}+ {1\over 8} \sum_{l=1}^4{1\over a_4 -z_l}
- {5 \over 4} {\partial \over \partial a_4 }\ln T\right), \quad \hbox{etc.}
\label{Eqeight}
\end{eqnarray}
In deriving eq. (\ref{Eqfour}) we have used the property of $T$ under the
$SL(2,C)$ linear fractional transformation. It gives the following identities:
\begin{equation}
\sum_{i=1}^6 a_i^n {\partial\over \partial a_i } \ln T = 
\left\{ 
\begin{array}{ll}
0, & n =0,\\
-3, & n=1,\\
-\sum_{i=1}^6 a_i, & n=3. 
\end{array}
\right.
\label{Eqidentity}
\end{equation}

Now we prove that all  terms containing $I_i(r)$ give total derivative
terms. The formula for $I_i(r)$ $(i=1,2,3)$ is already given in eq. (\ref{Eqnine}) and
the formula for $I_4(r)$ is
\begin{eqnarray}
&  & I_4(r) \prod_{l=1}^4 {(a_4-z_l)\over y(z_l) }
{1\over T^5 \prod_{k<l}^6 
a_{kl} }  
\nonumber\\
& & \quad = {1\over4}\sum_{i=1}^3 {\partial \over \partial a_i} 
\left\{ { 1\over {r -a_4} }{a_{i5}a_{i6}\over a_{45}a_{46}} 
{1\over  T^5 \prod_{k<l}^6 a_{kl}} \prod_{l=1}^4 {(a_4-z_l) \over y(z_l) }
\right\}  
\nonumber \\
& &  + {1\over 4}\sum_{l=1}^4  {\partial \over \partial z_l} 
\left\{ {1 \over  (r -a_4)} {(a_5-z_l)(a_6-z_l)\over a_{45}a_{46} }
{1\over  T^5 \prod_{k<l}^6 a_{kl} } \prod_{l=1}^4 {(a_4-z_l) \over y(z_l) }
\right\}  .
\label{Eqtena}
\end{eqnarray}
In the process of deriving (\ref{Eqtena}) we used (\ref{Eqidentity}) again to express the derivative
of $T$ with respect to $a_4$ in terms of the derivatives of $a_i$ ($i=1,2,3$):
\begin{equation}
{\partial \over \partial a_4 } \ln T = { 1\over a_{45}a_{46} } \left\{
2 (a_5+ a_6) - \sum_{i=1}^4 a_i - \sum_{i=1}^3 a_{i5}a_{i6} 
{\partial \over \partial a_i } \ln T \right\} .
\label{Eqeleven}
\end{equation}

\section *{Appendix B}
In this appendix we  give the complete 4-graviton amplitude, the kinematic factor and
collect some formulas used in the discussion of the Type II Superstrig
case. The vertex operator for the emission of a graviton with momentum $k^{\mu}$ and
polarization tensor $\epsilon^{\mu\nu}$ is
\begin{eqnarray}
V(k, \epsilon; z, \bar{z}) & =&  
\epsilon^{\mu\nu} :
(\partial_z X_{\mu}(z, \bar{z})  + i k\cdot \psi(z) \psi_{\mu}(z))
\nonumber \\
& & \times (\partial_{\bar{z}} X_{\nu}(z, \bar{z}) 
 + i k \cdot \tilde{\psi}(\bar{z}) \tilde{\psi}_{\nu}(\bar{z})
) \, e^{ i \cdot k X(z, \bar{z}) } : . 
\end{eqnarray}
The complete two-loop 4-graviton amplitude is
\begin{eqnarray}
& & AII(k_i,\epsilon_i) = c_{II} \, K\, 
\int { d^2 a_1 d^2 a_2 d^2 a_3 \,
|a_{45}\,a_{46}\,a_{56}|^2 \over 
T^5 \prod_{i<j}^6 |a_{ij}|^2 } 
\prod_{l=1}^4 {d^2 z_l (r-z_l)(\bar{s}-\bar{z}_l)
\over
|y(z_l)|^2 } 
\nonumber \\
& & \qquad \times \left\{ \left(I(r)\bar{I}(\bar{s}) + 
{5\over 4}\left( {\pi \over T \, y(r)\bar{y}(\bar{s})
} \int {d^2 w (r-w)
(\bar{s}-\bar{w})\over |y(w)|^2 }\right)^2 \right) \langle \prod e^{i k\cdot X}\rangle
\right.
\nonumber \\
& & \qquad \qquad + {1\over 16}  \left\langle \partial X(r+)\cdot \partial X(r-)
\bar{\partial} X(\bar{s}+) \cdot \bar{\partial} X(\bar{s}-) 
 \prod e^{i k\cdot X}\right\rangle  
\nonumber \\
& & \qquad \qquad \left. -
{1\over 16} \left\langle \partial X(r+)\cdot \partial X(r-)
\bar{\partial} X(\bar{s}+) \cdot \bar{\partial} X(\bar{s}-)\right\rangle   
\left\langle \prod e^{i k\cdot X}\right\rangle \right\} ,
\label{EqAIIcomplete}
\end{eqnarray}
where $c_{II}$ is an overall constant which should be fixed by unitarity and $K$ is 
the standard kinematic factor (for $\epsilon^{\mu\nu}_i = \epsilon^{\mu}_i
\tilde{\epsilon}^{\nu}_i$) \cite{GreenSchwarz,IengoZhu}:
\begin{eqnarray}
K & = & K_R \cdot K_L, 
\\
K_R & = & -{1\over 4} (st\epsilon_1\cdot\epsilon_3 \epsilon_2\cdot\epsilon_4
+su \epsilon_2\cdot\epsilon_3 \epsilon_1\cdot\epsilon_4
+ tu \epsilon_1\cdot\epsilon_2 \epsilon_3\cdot\epsilon_4 )
\nonumber \\
& & + {1\over 2} \, s \, ( \epsilon_1\cdot k_4 \epsilon_3\cdot k_2 \epsilon_2 \cdot
\epsilon_4  + \epsilon_2\cdot k_3 \epsilon_4\cdot k_1 \epsilon_1 \cdot
\epsilon_3  
\nonumber \\
& & \qquad + \epsilon_1\cdot k_3 \epsilon_4\cdot k_2 \epsilon_2 \cdot
\epsilon_3  + \epsilon_2\cdot k_4 \epsilon_3\cdot k_1 \epsilon_1 \cdot
\epsilon_4  )
\nonumber \\
& & + {1\over 2} \, t \, ( \epsilon_2\cdot k_1 \epsilon_4\cdot k_3 \epsilon_1 \cdot
\epsilon_3  + \epsilon_3\cdot k_4 \epsilon_1\cdot k_2 \epsilon_2 \cdot
\epsilon_4  
\nonumber \\
& & \qquad + \epsilon_2\cdot k_4 \epsilon_1\cdot k_3 \epsilon_3 \cdot
\epsilon_4  + \epsilon_3\cdot k_1 \epsilon_4\cdot k_2 \epsilon_1 \cdot
\epsilon_2  )
\nonumber \\
& & + {1\over 2} \, u \, ( \epsilon_1\cdot k_2 \epsilon_4\cdot k_3 \epsilon_2 \cdot
\epsilon_3  + \epsilon_3\cdot k_4 \epsilon_2\cdot k_1 \epsilon_1 \cdot
\epsilon_4  
\nonumber \\
& & \qquad + \epsilon_1\cdot k_4 \epsilon_2\cdot k_3 \epsilon_1 \cdot
\epsilon_4  + \epsilon_3\cdot k_2 \epsilon_4\cdot k_1 \epsilon_1 \cdot
\epsilon_2  ),
\\
K_L & = & K_R(\epsilon \to \tilde{\epsilon}).
\end{eqnarray}
Here $s$, $t$ and $u$ are the standard Mandelstam variables for the 4-gravitons. 
Defining $t_{\mu_1\nu_1\mu_2\nu_2\mu_3\nu_3\mu_4\nu_4}$ as follows:
\begin{equation}
K_R = t_{\mu_1\nu_1\mu_2\nu_2\mu_3\nu_3
 \mu_4\nu_4} \epsilon_1^{\mu_1} k_1^{\nu_1}
\epsilon_2^{\mu_2} k_2^{\nu_2}\epsilon_3^{\mu_3} k_3^{\nu_3}
\epsilon_4^{\mu_4} k_4^{\nu_4}, 
\end{equation}
the $R^4$ term is given as follows:
\begin{equation}
R^4 = t^{\mu_1\nu_1\mu_2\nu_2\mu_3\nu_3 \mu_4\nu_4} 
t^{\rho_1\sigma_1\rho_2\sigma_2\rho_3\sigma_3\rho_4\sigma_4}
R_{\mu_1\nu_1\rho_1\sigma_1}R_{\mu_2\nu_2\rho_2\sigma_2}
R_{\mu_3\nu_3\rho_3\sigma_3}R_{\mu_4\nu_4\rho_4\sigma_4}. 
\end{equation}

To prove the independence of the amplitude $AII$ in eq. (\ref{EqAIIcomplete}) 
we need the following identity:
\begin{eqnarray}
& & {1\over 4} \sum_{j=1}^6 {1\over \bar{A}_j - \bar{s} } \,
{\partial^2\over \partial \bar{A}_j
\partial A_i} \ln T
\nonumber \\
 & & \qquad =  { 1\over \prod_{j\neq i}^6 (A_i - A_j) }
\, \left( { \pi \over T \, \bar{Y}(\bar{s}) } \,  
\int { d^2 v (A_i -v)(\bar{s} - \bar{v}) \over | Y(v)|^2 } \right)^2,
\label{Eqtheaa}
\end{eqnarray}
which was proved in \cite{ZhuDr}. 
Notice also the following identity
\begin{eqnarray}
 { \prod_{l=1}^6 (r - Z_l) \over Y^2(r) } = 1 + \sum_{i=1}^6 {1 \over
r - A_i } \, { \prod_{l=1}^6 (A_i - Z_l) \over  \prod_{j\neq i}^6 (A_i-
A_j) }
\hskip 3cm & &
\nonumber \\
 \qquad =  { \prod_{l=1}^6 (x - Z_l) \over Y^2(x) } +
 \sum_{i=1}^6 \left( {1 \over r - A_i } - {1 \over
x - A_i }\right)  \, { \prod_{l=1}^6 (A_i - Z_l) \over
\prod_{j\neq i}^6 (A_i- A_j) },&  &
\end{eqnarray}
we have
\begin{eqnarray}
& & \left( {\pi \over T \, Y(r) \bar{Y}(\bar{s}) } \, \int
{ d^2 v (r -v)(\bar{s} - \bar{v}) \over | Y(v)|^2 } \right)^2
\, \prod_{l=1}^4 (r -Z_l)
\nonumber \\
& &  =  \left( {\pi \over T \, Y(x) \bar{Y}(\bar{s}) } \, \int
{ d^2 v (x-v)(\bar{s} - \bar{v}) \over | Y(v)|^2 } \right)^2
\, \prod_{l=1}^4 (x -Z_l)
\nonumber \\
& &  + \sum_{i=1}^6 \left( {1 \over r - A_i } - {1 \over
x - A_i }\right)  \, { \prod_{l=1}^4(A_i - Z_l) \over  \prod_{j\neq i}^6
(A_i- A_j) }  
\, \left( { \pi \over T \, \bar{Y}(\bar{s}) } \,
\int { d^2 v (A_i -v)(\bar{s} - \bar{v}) \over | Y(v)|^2 } \right)^2
\nonumber \\
& &  =  \left( {\pi \over T \, Y(x) \bar{Y}(\bar{s}) } \, \int
{ d^2 v (x-v)(\bar{s} - \bar{v}) \over | Y(v)|^2 } \right)^2
\, \prod_{l=1}^4 (x -Z_l)
\nonumber \\
& & + {1\over 4 }
\sum_{i=1}^6 \,\prod_{l=1}^4 (A_i-Z_l)\left( {1 \over r - A_i }
- {1 \over x - A_i }\right)  \, \sum_{j=1}^6 {1\over \bar{A}_j - \bar{s} } \,
{\partial^2\over \partial \bar{A}_j \partial A_i} \ln T,
\label{Eqaddb}
\end{eqnarray}
by making use of (\ref{Eqtheaa}).

By using eq. (\ref{Eqthreemm}) for both the left sector and right sector
and also the above equation with $r \to S_A(r)$ we have
\begin{eqnarray}
& &\left\{  I(S_A(r)) \bar{I}(\bar{S}_A(\bar{s}) ) + {5\over 4}
\left( { \pi \over T} \,
\int {d^2 v (S_A(r)-v)
(\bar{S}_A(\bar{s})-\bar{v})\over {Y(S_A(r))\bar{Y}(\bar{S}_A(\bar{s}))}
 |y(v)|^2 }\right)^2 \right\}
\nonumber \\
& & \qquad \times
 \prod_{k=1}^4  { (S_A(r) - Z_k) (\bar{S}_A(\bar{s}) - \bar{Z}_k)
 \over |Y(Z_k)|^2 } \,
{ 1\over T^5 \prod_{i<j=1}^6 |A_{ij}|^2 }
\nonumber \\
& & \quad  = \left\{
I_M(x) \, \bar{I}_M(\bar{w})
 + {5\over 4} \left( { \pi \over T\, Y(x) \bar{Y}(\bar{w}) } \int {d^2 v (x-v)
(\bar{w}-\bar{v})\over |Y(v)|^2 }\right)^2 \right\}
\nonumber \\
& & \qquad \qquad \times \prod_{k=1}^4  { (x - Z_k)(\bar{w} - \bar{Z}_k)
\over |Y(Z_k)|^2 } \,
{ 1\over T^5 \prod_{i<j=1}^6 |A_{ij}|^2 }
\nonumber \\
& &  \qquad - { 1\over 4}\, \sum_{i=1}^6{\partial \over \partial A_i }
\left[ \left( { 1\over A_i - S_A(r) } - {1\over  A_i - x} \right)
\right.
\nonumber \\
& & \qquad \qquad \times \left. \prod_{l=1}^4
{ (A_i - Z_l)(\bar S_A(\bar s)- \bar{Z}_l) \over |Y(Z_l)|^2  } \,
{ \bar I(\bar S_A(\bar s))
\over T^5 \prod_{j<k=1}^6 |A_{jk}|^2  } \right]
\nonumber \\
& &  \qquad - { 1\over 4}\,\,
 \sum_{i=1}^6{\partial \over \partial \bar{A}_i }
\left[ \left( { 1\over \bar{A}_i - \bar{S}_A(\bar{s}) } - {1\over  \bar{A}_i
-\bar{w}} \right)
\right.
\nonumber \\
& & \qquad \qquad \times \left. \prod_{l=1}^4
{ (\bar{A}_i - \bar{Z}_l)(x- {Z}_l) \over |Y(Z_l)|^2  } \,
{ {I}_M({x}) \over T^5 \prod_{j<k=1}^6 |A_{jk}|^2  } \right] .
\label{finalAppendix}
\end{eqnarray}
In deriving the above equation we have used the fact that the only
non-holomorphic (meromorphic) part in both $I(x)$ and $I_M(x)$ is from the
terms $ -{5\over 4} \sum_{i=1}^6 {1\over x- A_i}{\partial\over \partial A_i}
\ln T$, i.e. $T$ depends on both $A_i$ and $\bar{A}_i$.

\end{document}